\RequirePackage{fix-cm}
\documentclass[twocolumn,epjc3]{svjour3}
\smartqed  
\RequirePackage{graphicx}
\pdfoutput=1
\usepackage{amsmath,amssymb,graphicx}
\graphicspath{{figs/}}
\usepackage{epsf,verbatim}
\usepackage{hyperref}
\usepackage{comment}
\usepackage{color}
\usepackage{slashed}
\usepackage{subfigure}
\usepackage[usenames,dvipsnames]{xcolor}
\usepackage{comment}
\usepackage{mdwlist, paralist}
\usepackage{rotating}
\usepackage{multirow}

\usepackage[utf8]{inputenc}

\newcommand{\Br}{{\mathrm{Br}}}

\newcommand{\ord}{\mathcal{O}}
\newcommand{\TeV}{~{\mbox{TeV}}}
\newcommand{\abi}{~{\mbox{ab}^{-1}}}
\newcommand{\fbi}{~{\mbox{fb}^{-1}}}
\newcommand{\fb}{~{\mbox{fb}}}
\newcommand{\GeV}{~{\mbox{GeV}}}

\newcommand{\bb}[1]{\raisebox{-1.5ex}[0pt][0pt]{\shortstack{#1}}}

\journalname{Eur. Phys. J. C}
\begin{document}

\title{Exotic Higgs Decay $h\to\phi\phi\to 4b$ at the LHeC
}


\author{Shang Liu\thanksref{e1,addr1}
        \and
        Yi-Lei Tang\thanksref{e2,addr2} 
        \and
        Chen Zhang\thanksref{e3,addr1}
        \and
        Shou-hua Zhu\thanksref{e4,addr1,addr2,addr3}
}

\thankstext{e1}{e-mail: liushang@pku.edu.cn}
\thankstext{e2}{e-mail: tangyilei15@pku.edu.cn}
\thankstext{e3}{e-mail: larry@pku.edu.cn}
\thankstext{e4}{e-mail: shzhu@pku.edu.cn}


\institute{Institute of Theoretical Physics $\&$ State Key Laboratory of Nuclear Physics and Technology, Peking University, Beijing 100871, China \label{addr1}
           \and
           Center for High Energy Physics, Peking University, Beijing 100871, China \label{addr2}
           \and
           Collaborative Innovation Center of Quantum Matter, Beijing 100871, China \label{addr3}
}

\date{Received: date / Accepted: date}

\maketitle

\begin{abstract}
We study the exotic decay of the 125 GeV Higgs boson ($h$) into a pair of light
spin-0 particles ($\phi$) which subsequently decay and result in a $4b$
final state. This channel is well motivated in models with an extended
Higgs sector. Instead of searching at the Large Hadron Collider (LHC) and the High
Luminosity LHC (HL-LHC) which are beset by large Standard
Model (SM) backgrounds, we investigate this decay channel at the much
cleaner Large Hadron Electron Collider (LHeC). With some simple selection
cuts this channel becomes nearly free of background at this $ep$ machine,
in sharp contrast to the situation at the (HL-)LHC. With a parton level
analysis we show that for the $\phi$ mass range $[20,60]\GeV$, with
$100\fbi$ luminosity the LHeC is generally capable of constraining
$C_{4b}^2\equiv\kappa_{V}^2\times
\Br(h\rightarrow\phi\phi)\times\Br^2(\phi\rightarrow b\bar b)$
($\kappa_{V}$ denotes the $hVV(V=W,Z)$ coupling strength relative to the
SM value) to a few percent level ($95\%$ CLs). With $1\abi$ luminosity
$C_{4b}^2$ at a few per mille level can be probed. These sensitivities
are much better than the HL-LHC performance and demonstrate the important
role expected to be played by the LHeC in probing exotic Higgs decay
processes, in addition to the already proposed invisible Higgs decay
channel.
\end{abstract}

\section{Introduction}
\label{intro}
The discovery of the 125 GeV Higgs boson (denoted as $h$)
~\cite{Chatrchyan:2012ufa,Aad:2012tfa} not only deepens our understanding
of the mechanism of electroweak symmetry breaking but also opens new
avenues for searching for physics beyond the Standard Model (SM) which is
required to clarify the unexplained theoretical and observational issues
such as the problem of naturalness, the existence of dark matter and the
observed baryon asymmetry of the universe. One of such avenues is exotic
Higgs decay~\footnote{In this paper we study exotic decays of the
$125\GeV$ Higgs boson. For exotic decays of additional Higgs bosons we
refer interested readers to a recent study~\cite{Kling:2016opi}.}, which
is only loosely constrained by Higgs signal strength
measurements. The combination of ATLAS and CMS Run I results constrains
undetected Higgs decay branching ratio to be smaller than about 34\% at
95\% C.L. assuming $\kappa_V\leq1$~\cite{Khachatryan:2016vau}($\kappa_V$
denotes $hVV(V=W,Z)$ coupling strength relative to SM
assuming $\kappa_V\equiv\kappa_W=\kappa_Z$). The ultimate sensitivity
on undetected Higgs decay branching ratio via indirect measurements at
the High Luminosity Large Hadron Collider (HL-LHC) is estimated to be
$\ord (5-10\%)$~\cite{Curtin:2013fra}. On the other hand, due to the
expected extremely narrow width of the Higgs boson, even a rather weak
coupling between it and any new light degrees of freedom can naturally
induce a sizable exotic decay branching fraction. One such possibility
is $h \to \phi\phi$, where $\phi$ denotes a light spin-0 particle,
with mass less than about 62.5 GeV so that this decay channel is
kinematically allowed. $\phi$ can be CP-even or CP-odd, or even a
CP-mixed state. If its mass is greater than $2m_{b}$, then in most
models which approximately obey Yukawa ordering $\phi$ will mainly
decay to $b\bar{b}$. This decay channel is well motivated in
a wide class of Beyond the Standard Model (BSM)
theories~\cite{Curtin:2013fra}, such as the Next to Minimal
Supersymmetric Standard Model (NMSSM), Higgs singlet extension of the
SM, general extended Higgs sector models~\cite{Mao:2016jor}, and Little Higgs
Models. Quite a few phenomenology studies already exist with respect
to this channel at the LHC~\cite{Cao:2013gba,Cheung:2007sva,Carena:2007jk,Kaplan:2011vf,Kaplan:2009qt}, with or without using jet
substructure techniques. Due to large QCD backgrounds in gluon fusion
and vector boson fusion channels, the LHC searches generally focus
on the VH associated production channel. However this channel suffers
from large top quark backgrounds. A recent ATLAS
analysis~\cite{Aaboud:2016oyb} using $3.2\fbi$ $13\TeV$ data made the
first attempt to constrain this channel using WH associated production
but the sensitivity is currently quite weak (even
$\Br(h\to\phi\phi\to 4b)=100\%$ cannot be constrained assuming
$\kappa_V=1$).

The not-so-clean hadron-hadron collision environment motivates us to
consider better places to search for this exotic Higgs decay channel.
Here we consider using the Large Hadron Electron Collider
(LHeC)~\cite{AbelleiraFernandez:2012cc} to explore
$h\to\phi\phi\to 4b$. The LHeC is a proposed lepton-hadron collider
which is designed to collide a $60\GeV$ electron beam with the
$7\TeV$ proton beam of the HL-LHC. It is supposed to run
synchronously with the HL-LHC and may deliver an integrated luminosity
as high as $1000\fbi$~\cite{Bruening:2013bga}. The electron beam
may have $-0.9$ polarization~\cite{Bruening:2013bga}. It is worth
noticing that with such high collision energy and luminosity, the
LHeC indeed becomes a Higgs boson factory~\cite{Bruening:2013bga}.
With Higgs boson production cross section of about $200\fbi$, the
LHeC will provide amazing opportunities for precision Higgs physics,
due to the fact that major QCD backgrounds will be much smaller
than LHC and the complication due to pile-up will be greatly
reduced. Previous studies on Higgs physics at the LHeC include
measuring bottom Yukawa coupling~\cite{Jarlskog:1990dv,Han:2009pe,AbelleiraFernandez:2012cc}, anomalous gauge-Higgs
coupling~\cite{Biswal:2012mp,Cakir:2013bxa,Senol:2012fc}, invisible
Higgs decay~\cite{Tang:2015uha} and MSSM Higgs
production~\cite{Zhe:2011yr}. Studies on charm Yukawa measurements
has been reported in ~\cite{Klein:2015hcc}. The impact of double
Higgs production at the higher energy $ep$ collider FCC-he on
Higgs-self coupling measurement has also been
studied~\cite{Kumar:2015tua,Kumar:2015kca}.

To quantitatively estimate the sensitivity of the LHeC to the
exotic Higgs decay $h\to\phi\phi\to 4b$, we perform a parton level
study for the signal and background in the next section. The signal
definition depends on the required number of $b$-tagged jets. Here
for simplicity and a clear identification of signal we require
tagging at least 4 $b-$tagged jets. We provide the expected LHeC
sensitivity for $\phi$ mass between $15\GeV$ and $60\GeV$ and
investigate the robustness of our results under variation of
$b$-tagging performance and pseudorapidity coverage. We also
translate our results into the expected exclusion power in the
parameter space of the Higgs singlet extension of the SM. In the
last section we present our discussion and conclusion.

\section{Collider Sensitivity}
\label{sec:cs}
The exotic Higgs dacay $h\to\phi\phi\to 4b$ can be simply characterized
by the following effective interaction Lagrangian for a new real scalar
degree of freedom $\phi$,
\begin{align}
\mathcal{L}_{eff}=\lambda_{h}vh\phi^2+\lambda_{b}\phi\bar b b
+\mathcal{L}_{\phi\,\text{decay,other}}
\label{eqn:Leff}
\end{align}
In the above $v=246\GeV$. $\lambda_{h}$ and $\lambda_{b}$ are real
dimensionless parameters and $\mathcal{L}_{\phi\,\text{decay,other}}$
denotes the part of Lagrangian which mediates the decay of $\phi$ into
final states other than $b\bar b$. The part of Lagrangian
$\mathcal{L}_{eff}-\mathcal{L}_{\phi\,\text{decay,other}}$ has been taken as
CP-even without loss of generality. New physics may also modify
$hVV (V=W,Z)$ coupling which affects the Higgs production rate and
kinematics. We assume the $hVV (V=W,Z)$ coupling is purely CP-even.
Assuming narrow width approximation is valid for both $h$ and $\phi$,
we can express the collider reach for $h\to\phi\phi\to 4b$ via the
following quantity
\begin{align}
C_{4b}^2=\kappa_{V}^2\times\Br(h\rightarrow\phi\phi)\times\Br^2
(\phi\rightarrow b\bar b)
\label{eqn:c4b2}
\end{align}
for a given value of the $\phi$ mass $m_\phi$.

There are two major Higgs production channels at the LHeC: charged
current (CC) and neutral current (NC). Due to the accidentally suppressed
electron NC coupling, NC Higgs cross section is much less than that of
CC~\cite{Han:2009pe}. Therefore in the following we only focus on CC
process, although in a more detailed analysis the NC process should also
be included to enhance the overall statistical significance.

\begin{figure}[ht]
\begin{center}
\includegraphics[width=0.4\textwidth]{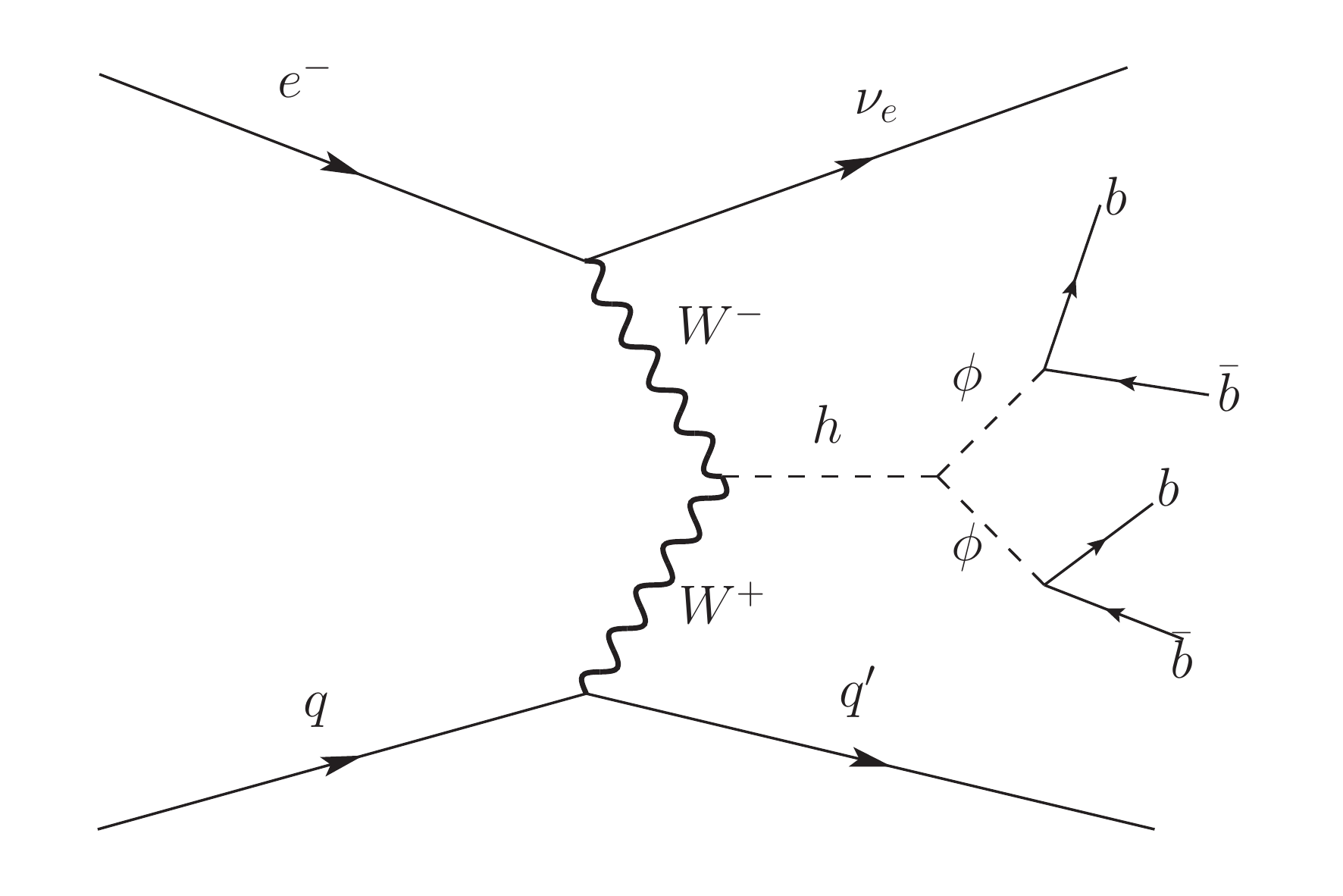}
\caption{Feynman diagram of the CC signal process.}
\label{fig:cc_sig}
\end{center}
\end{figure}

The signal process of CC Higgs production is
\begin{align}
eq\to\nu_{e}hq'\to\nu_{e}\phi\phi q'\to\nu_{e}b\bar bb\bar bq'
\end{align}
The corresponding Feynman diagram is shown in Fig.~\ref{fig:cc_sig}.
The signal signature thus contains at least 5 jets (in which
at least 4 jets are $b$-tagged) plus missing transverse energy.

\begin{figure*}[ht]
\begin{center}
\includegraphics[width=2.0in]{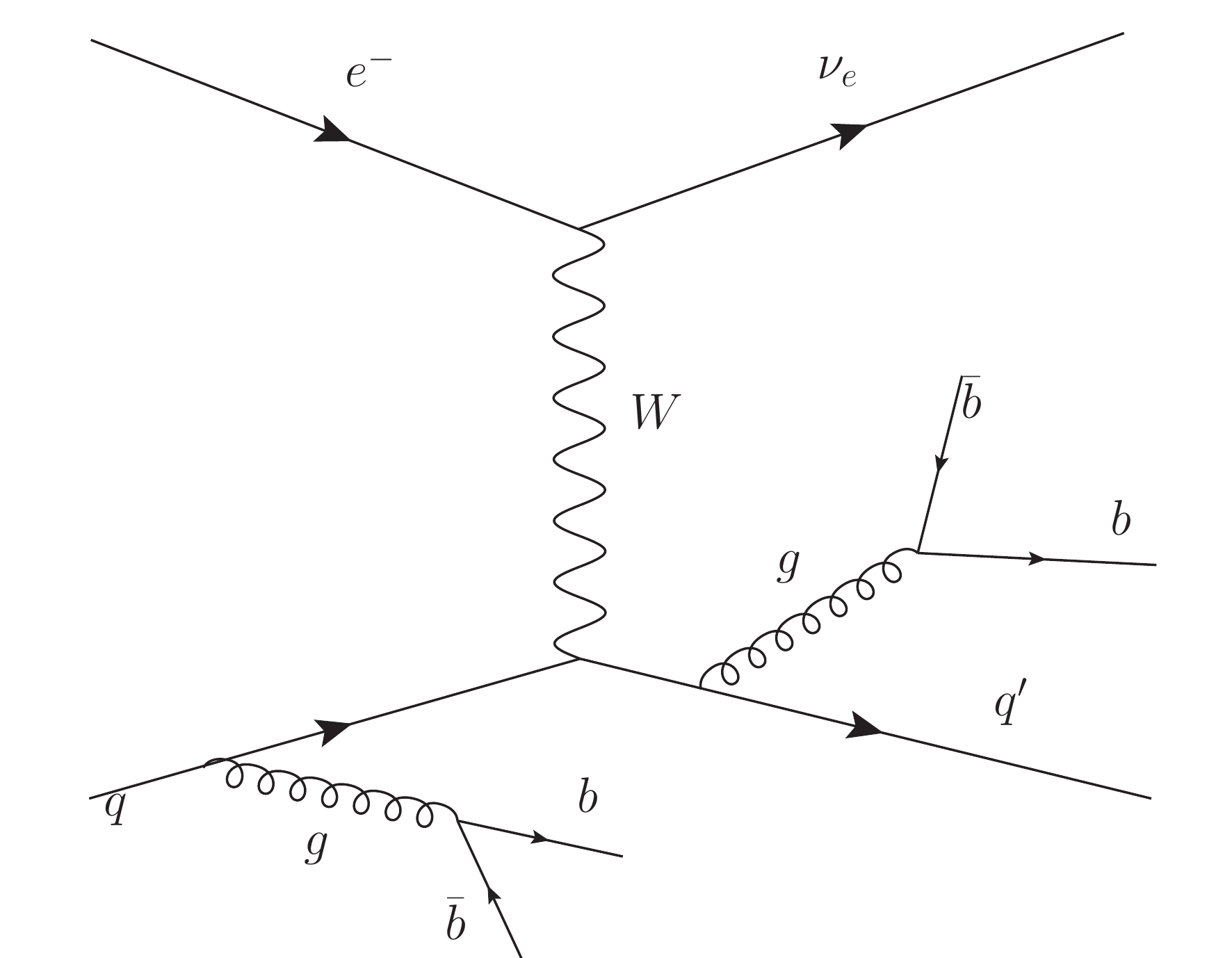}\hspace{20pt}
\includegraphics[width=2.0in]{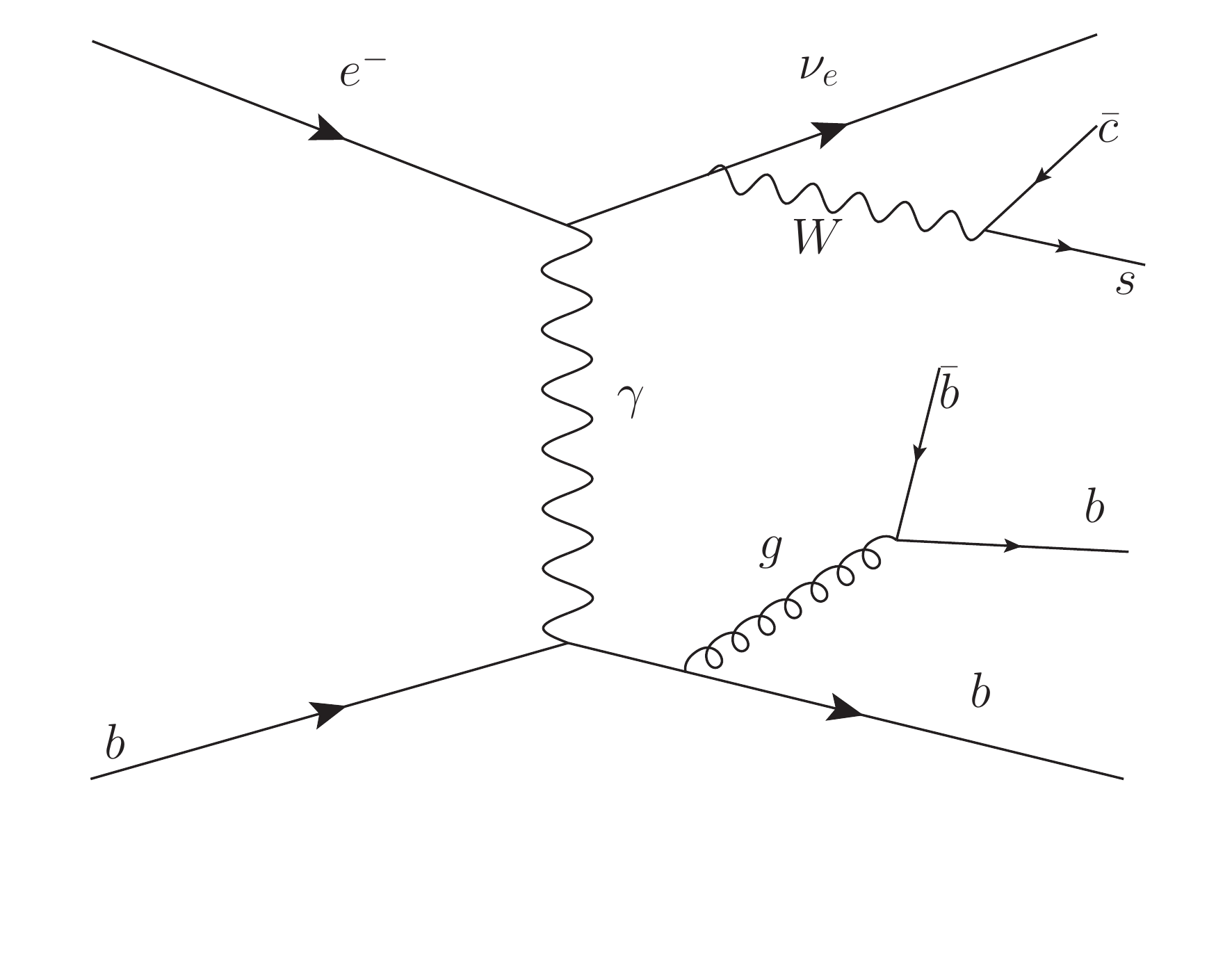}  \\
\end{center}
\includegraphics[width=2.0in]{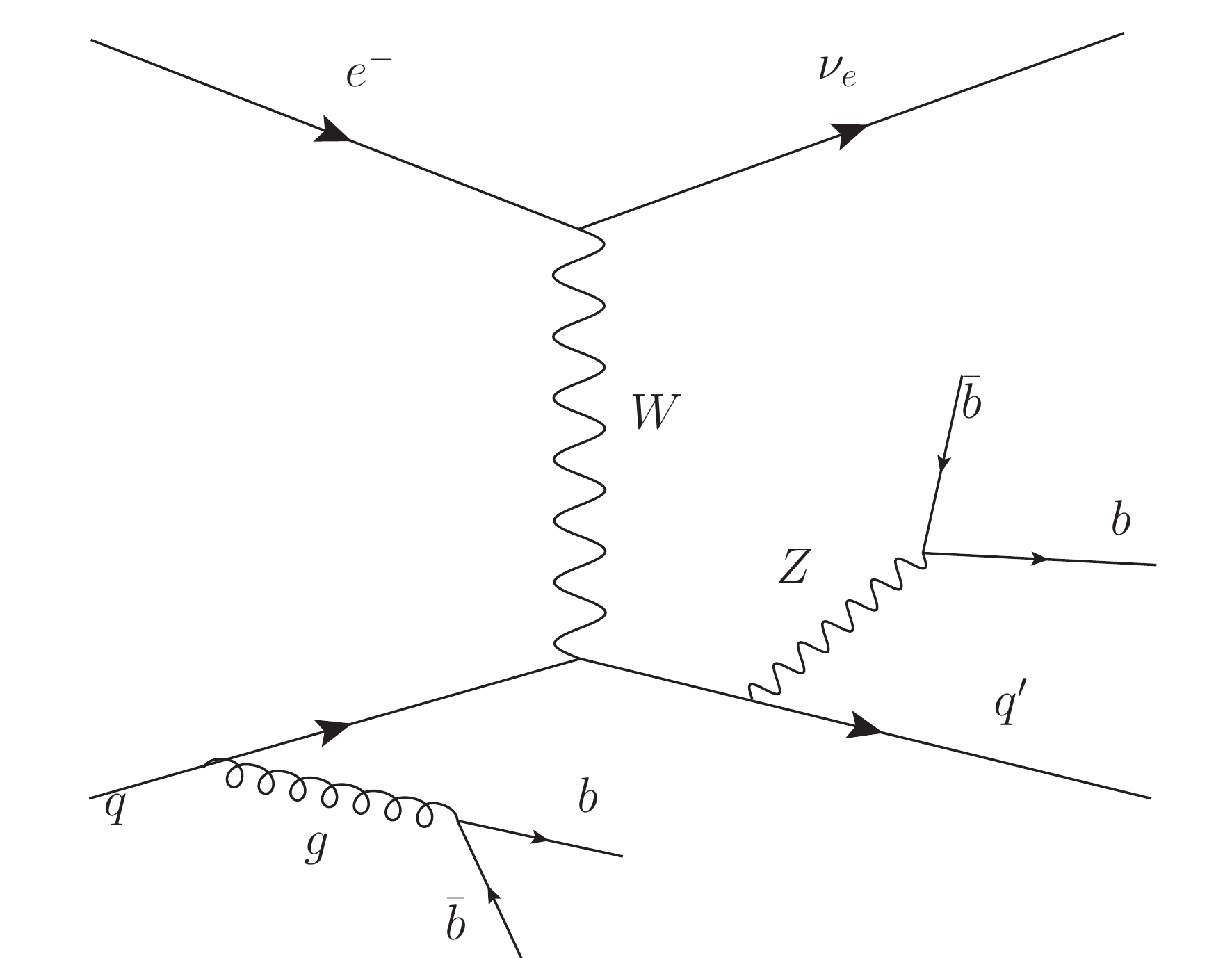}\hspace{25pt}
\includegraphics[width=2.0in]{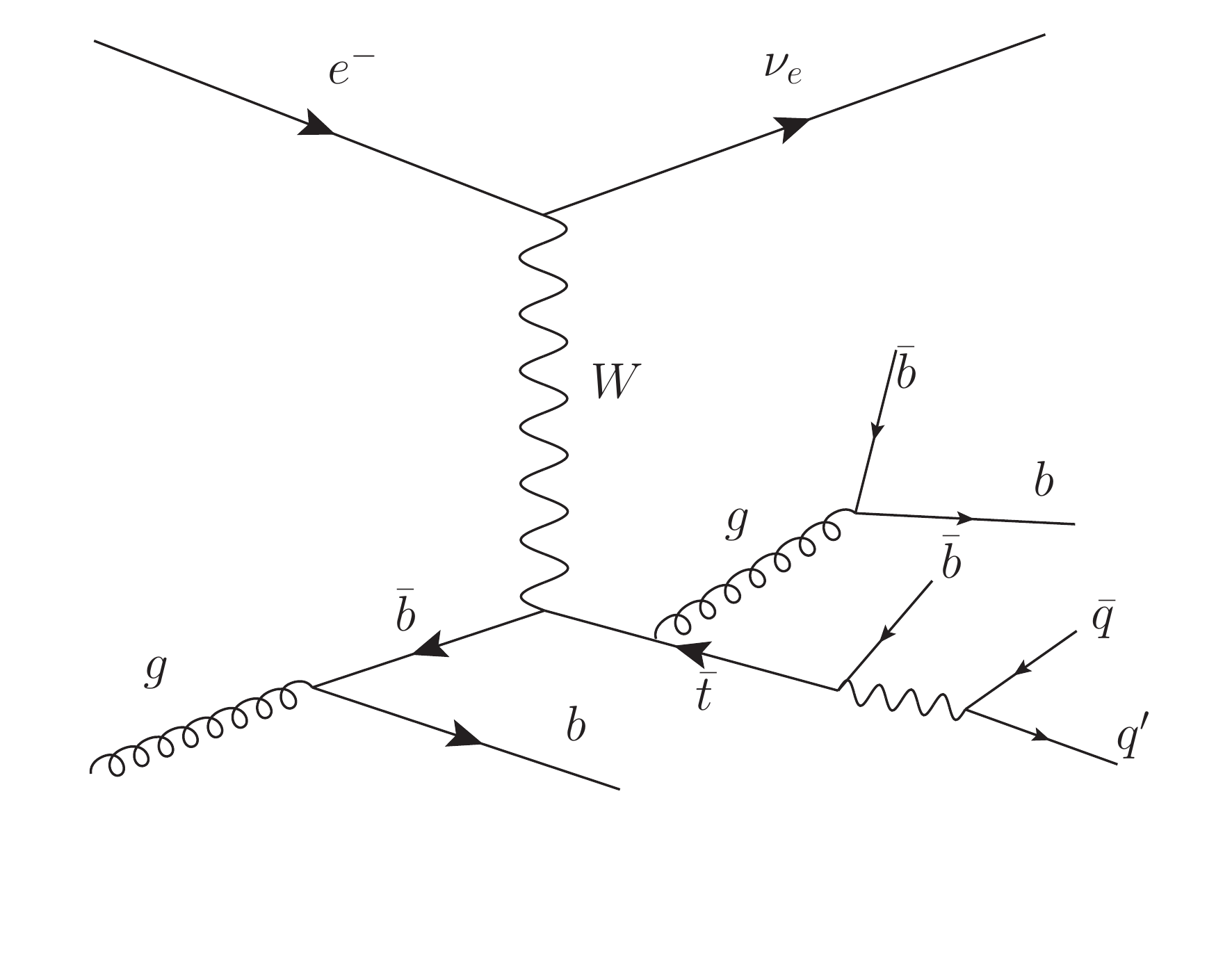}\hspace{10pt}
\includegraphics[scale=0.3]{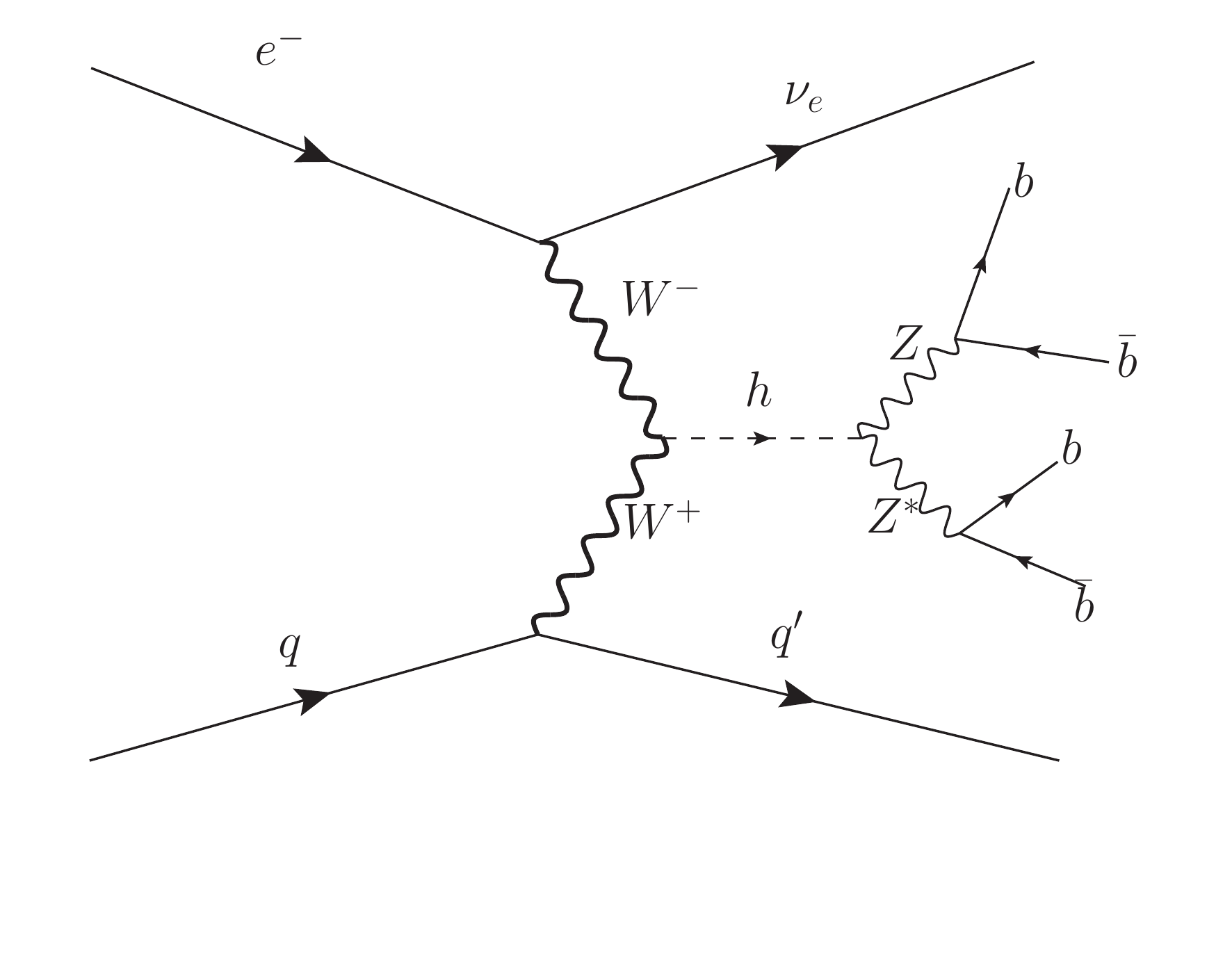}
\caption{\label{fig:bkggraph}Representative Feynman diagrams of
the following backgrounds: CC multijet (top left), CC $W+$jets
(top right), CC $Z+$ jets (bottom left), CC $t+$jets
(bottom center), CC $h+$jets (bottom right).}
\end{figure*}

The backgrounds can be classified into charged current (CC) deeply
inelastic scattering (DIS) backgrounds and photoproduction (PHP)
backgrounds. From a parton level point of view, CC DIS backgrounds
all have genuine $\slashed{E}_T$ in the final state which comes
from neutrinos produced in the hard scattering or decay of
heavy resonances ($W,Z,t,h$). When it comes to PHP backgrounds,
only PHP production of heavy resonances ($W,Z,t,h$) could produce
such genuine $\slashed{E}_T$. However, these PHP processes (which
involve the on-shell production of heavy resonances ($W,Z,t,h$))
are found to be negligible in the total background. On the other
hand, PHP multijet production (including heavy flavor jets) could
produce $\slashed{E}_T$ only via energy mismeasurement and
neutrinos from hadron decay, which means that it could be
suppressed efficiently through a sufficiently large requirement
of $\slashed{E}_T$.

The CC backgrounds can be further classified according to the
number of heavy resonances ($W,Z,t,h$) produced which further
decay to result in a large number of $b$-tagged jets. We found that
if in one process the number of heavy resonances involved is greater
than or equal to two, then its contribution to the total background
is always negligible. Therefore in the following we only consider
the following CC backgrounds: CC multijet, CC $W+$jets, CC $Z+$jets,
CC $t+$jets, CC $h+$jets. Here "multijet" and "jets" contain jets
of all flavor ($g,u,d,s,c,b$). Higgs decay to $4b$ via SM processes
is also included as a background, in CC $h+$ jets.
Fig.~\ref{fig:bkggraph} displays representative Feynman diagrams
for these background processes.

To simulate the signal and backgrounds, we implement the effective
interaction in Eq.~\eqref{eqn:Leff} into
FeynRules~\cite{Alloul:2013bka}. The generated model file together
with the SM is then imported by MadGraph5\_aMC@NLO~\cite{Alwall:2014hca}. The
Higgs boson mass is taken to be $m_h=125\GeV$. The $\phi$ mass is scanned
in the region $[15,60]\GeV$ with $1\GeV$ step size. The collider
parameter is taken to be $E_e=60\GeV, E_p=7\TeV$ with electron beam being
$-0.9$ polarized. The signal and background samples are generated by
MadGraph5\_aMC@NLO at leading order with NNPDF2.3 LO
PDF~\cite{Ball:2013hta} and the renormalization and factorization scale
is set dynamically by MadGraph default. The NLO QCD correction to the
signal process are known to be small~\cite{Jager:2010zm}. In the
following we take all the signal and background K-factors to be 1
although we expect the correct background normalization could be obtained
from data. We apply jet energy smearing according to the following
energy resolution formula
\begin{align}
\frac{\sigma_E}{E}=\frac{\alpha}{\sqrt{E}}\oplus\beta
\end{align}
where $\alpha=0.45\GeV^{1/2},\beta=0.03$~\cite{AbelleiraFernandez:2012cc}.
We consider the following four scenarios of $b-$tagging performance
for jets with $p_T>20\GeV$: ($\epsilon_b$
denotes the efficiency of $b-$jet, while $\epsilon_c$ and
$\epsilon_{g,u,d,s}$ denote the faking probability of $c-$jet and
$g,u,d,s-$jet respectively)
\begin{enumerate}[(A)]
\item $\epsilon_b=70\%,\epsilon_c=10\%,\epsilon_{g,u,d,s}=1\%$ \label{eqn:ba}
\item $\epsilon_b=70\%,\epsilon_c=20\%,\epsilon_{g,u,d,s}=1\%$ \label{eqn:bb}
\item $\epsilon_b=60\%,\epsilon_c=10\%,\epsilon_{g,u,d,s}=1\%$ \label{eqn:bc}
\item $\epsilon_b=60\%,\epsilon_c=20\%,\epsilon_{g,u,d,s}=1\%$ \label{eqn:bd}
\end{enumerate}
The LHeC detector (including the tracker) is expected to have a very
large pseudorapidity coverage~\cite{Gaddi:2015det} and therefore we assume
the $b-$tagging performance listed above is valid up to $|\eta|<5$. We will
also show the expected sensitivities with smaller $b-$tagging pseudorapidity
coverage $|\eta|<4$ and $|\eta|<3$ which turn out to change only slightly
compared to the $|\eta|<5$ case. Event analysis is performed by
MadAnalysis 5~\cite{Conte:2012fm}.

\begin{table*}[t!]
\caption{The cross section (in unit of fb) of the signal and major backgrounds after application of each cut in the corresponding row.
Lepton veto and electron anti-tagging is implicit in basic cuts. Signal corresponds to $C_{4b}^2=1,m_\phi=20\GeV$. Here we assume $b-$tagging performance
scenario \eqref{eqn:ba} and a $b-$tagging pseudorapidity coverage $|\eta|<5.0$. $E_0=40\GeV$ is assumed except that in the last row for
the signal and total background we show in parentheses the values corresponding to $E_0=60\GeV$.
\label{table:cutflow20}}
\begin{center}
\begin{tabular}{cccccccc}
\hline
\bb{Cross Section (fb)} & \bb{Signal} & \bb{Total Background} & CC & CC & CC & CC & CC \\
 & & & Multijet & $h+$jets & $t+$jets & $Z+$jets & $W+$jets \\
\hline
Basic cuts~\eqref{eqn:bcut} & 7.64 & $-$ & 357 & 0.81 & 118 & 14.9 & 26.6 \\
\hline
$\slashed{E}_T>E_0$~\eqref{eqn:met} & 4.32 & $-$  & 258 & 0.46 & 66.4 & 12.7 & 21.1 \\
\hline
$b-$tagging~\eqref{eqn:4btag} & 1.06 & $-$ & 8.5E-03 & 3.3E-03 & 8.1E-02 & 1.0E-02 & 2.2E-03   \\
\hline
$4b$ mass window~\eqref{eqn:m4b} & 1.04 & $-$ & 1.3E-03 & 1.7E-03 & 3.1E-03 & 2.6E-04 & 6.0E-05  \\
\hline
$2b$ mass window~\eqref{eqn:m2b} & 0.98 (0.58) & 3.2E-04 (2.6E-04) & 1.2E-04 & 7.3E-05 & 1.2E-04 & 4.1E-06 & 2.3E-06 \\
\hline
\end{tabular}
\end{center}
\end{table*}

\begin{table*}[t!]
\caption{The cross section (in unit of fb) of the signal and major backgrounds after application of each cut in the corresponding row.
Lepton veto and electron anti-tagging is implicit in basic cuts. Signal corresponds to $C_{4b}^2=1,m_\phi=40\GeV$. Here we assume $b-$tagging performance
scenario \eqref{eqn:ba} and a $b-$tagging pseudorapidity coverage $|\eta|<5.0$. $E_0=40\GeV$ is assumed except that in the last row for
the signal and total background we show in parentheses the values corresponding to $E_0=60\GeV$.
\label{table:cutflow40}}
\begin{center}
\begin{tabular}{cccccccc}
\hline
\bb{Cross Section (fb)} & \bb{Signal} & \bb{Total Background} & CC & CC & CC & CC & CC \\
 & & & Multijet & $h+$jets & $t+$jets & $Z+$jets & $W+$jets \\
\hline
Basic cuts~\eqref{eqn:bcut} & 11.3 & $-$ & 362 & 0.81 & 119 & 14.9 & 26.6 \\
\hline
$\slashed{E}_T>E_0$~\eqref{eqn:met} & 7.26 & $-$  & 263 & 0.46 & 67.6 & 12.7 & 21.1 \\
\hline
$b-$tagging~\eqref{eqn:4btag} & 1.78 & $-$ & 8.5E-03 & 3.3E-03 & 8.1E-02 & 1.0E-02 & 2.2E-03   \\
\hline
$4b$ mass window~\eqref{eqn:m4b} & 1.75 & $-$ & 1.3E-03 & 1.7E-03 & 3.2E-03 & 2.4E-04 & 5.9E-05  \\
\hline
$2b$ mass window~\eqref{eqn:m2b} & 1.62 (1.13) & 1.4E-03 (1.0E-03) & 2.5E-04 & 4.3E-04 & 6.1E-04 & 6.0E-05 & 1.0E-05 \\
\hline
\end{tabular}
\end{center}
\end{table*}

\begin{table*}[t!]
\caption{The cross section (in unit of fb) of the signal and major backgrounds after application of each cut in the corresponding row.
Lepton veto and electron anti-tagging is implicit in basic cuts. Signal corresponds to $C_{4b}^2=1,m_\phi=60\GeV$. Here we assume $b-$tagging performance
scenario \eqref{eqn:ba} and a $b-$tagging pseudorapidity coverage $|\eta|<5.0$. $E_0=40\GeV$ is assumed except that in the last row for
the signal and total background we show in parentheses the values corresponding to $E_0=60\GeV$.
\label{table:cutflow60}}
\begin{center}
\begin{tabular}{cccccccc}
\hline
\bb{Cross Section (fb)} & \bb{Signal} & \bb{Total Background} & CC & CC & CC & CC & CC \\
 & & & Multijet & $h+$jets & $t+$jets & $Z+$jets & $W+$jets \\
\hline
Basic cuts~\eqref{eqn:bcut} & 29.7 & $-$ & 358 & 0.81 & 119 & 15.0 & 26.6 \\
\hline
$\slashed{E}_T>E_0$~\eqref{eqn:met} & 17.5 & $-$  & 261 & 0.46 & 66.7 & 12.8 & 21.1 \\
\hline
$b-$tagging~\eqref{eqn:4btag} & 4.30 & $-$ & 8.6E-03 & 3.3E-03 & 8.1E-02 & 1.0E-02 & 2.2E-03   \\
\hline
$4b$ mass window~\eqref{eqn:m4b} & 4.22 & $-$ & 1.3E-03 & 1.7E-03 & 3.2E-03 & 2.5E-04 & 5.9E-05  \\
\hline
$2b$ mass window~\eqref{eqn:m2b} & 3.67 (2.28) & 1.4E-03 (9.4E-04) & 3.2E-04 & 3.0E-04 & 7.0E-04 & 4.3E-05 & 1.6E-05 \\
\hline
\end{tabular}
\end{center}
\end{table*}

The event selection in the $4b$-tagging case first requires at
least five jets satisfying the following basic cuts:
\begin{align}
p_{Tj}>20\GeV, |\eta_j|<5.0, \Delta R_{jj}>0.4
\label{eqn:bcut}
\end{align}
Events with additional charged leptons are vetoed. To suppress the
photoproduction background, we exclude events which can be tagged
by an electron tagger, and also require:
\begin{align}
\slashed{E}_T>E_0
\label{eqn:met}
\end{align}
Here $E_0$ denotes the threshold of transverse missing energy. In the
following we take $E_0=40\GeV$ as the default choice and assume PHP
backgrounds can be accordingly suppressed to a negligible level
compared to the total background. This rough estimate of missing
energy threshold is inspired by a naive simulation of direct
photoproduction $j+4b$ process.
~\footnote{According to previous experience~\cite{Klasen:1998cw}, in
PHP multijet processes the resolved component becomes smaller than
the direct component when a hard scale is involved. Therefore as
in ~\cite{Han:2009pe}, we do not expect resolved photoproduction
$j+4b$ to be a leading component in PHP backgrounds. For direct
photoproduction $j+4b$ (photon virtuality $Q^2<1\GeV^2$), we find a
cross section of about $0.9\fb$ after basic cuts and $4b-$tagging
requirement, with electron tagging and $4b$ and $2b$ invariant
mass requirement a cross section reduction by two orders of
magnitude could be expected. Because the total CC backgrounds are
at $10^{-3}\fb$ to a few times $10^{-4}\fb$ level depending on
$m_\phi$, the PHP backgrounds would become negligible if the
$\slashed{E}_T>E_0$ cut and perhaps missing energy isolation cuts
could bring down the PHP cross section by another two or three
orders of magnitude, which could be achieved for
$E_{0}\sim~40-60\GeV$ by our current rough estimation, given the
situation that the LHeC detector is supposed to have better
resolution and coverage than LHC.}
A thorough and detailed detector simulation of multijet
photoproduction would be needed to determine the best $E_0$ (perhaps
in synergy with appropriate missing energy isolation cuts or a cut
on the ratio $V_{ap}/V_p$ of transverse energy flow anti-parallel
and parallel to the hadronic final state transverse momentum
vector~\cite{Adloff:1999ah}), which is however beyond the scope of
the present paper. In the cut flow tables below we will also show
the signal and total background in the case that $E_0$ need be
increased to $60\GeV$.

Then we impose the $4b$-tagging requirement:
\begin{align}
\text{At least 4 }b\text{-tagged jets in }|\eta|<5.0
\label{eqn:4btag}
\end{align}
The 4 $b$-tagged jets which have the closest invariant mass to $m_h$
are required to have their invariant mass $m_{4b}$ lie in the following
mass window:
\begin{align}
|m_{4b}-m_h|<20\GeV
\label{eqn:m4b}
\end{align}
Finally we utilize the event structure of the signal: for the 4
$b-$tagged jets picked out in the previous step, we group them into two
pairs such that the absolute value of the invariant mass difference
between these two pairs is smallest among all grouping possibilities.
Then we require the invariant masses of these $b-$jet pairs both lie
in the following mass window:
\begin{align}
|m_{2b,i}-m_\phi|<10\GeV,\,i=1,2
\label{eqn:m2b}
\end{align}
Here $m_{2b,i},i=1,2$ denote the invariant mass of the two
"correctly" grouped $b-$jet pairs, respectively.

\begin{figure*}[ht]
\includegraphics[width=3.5in]{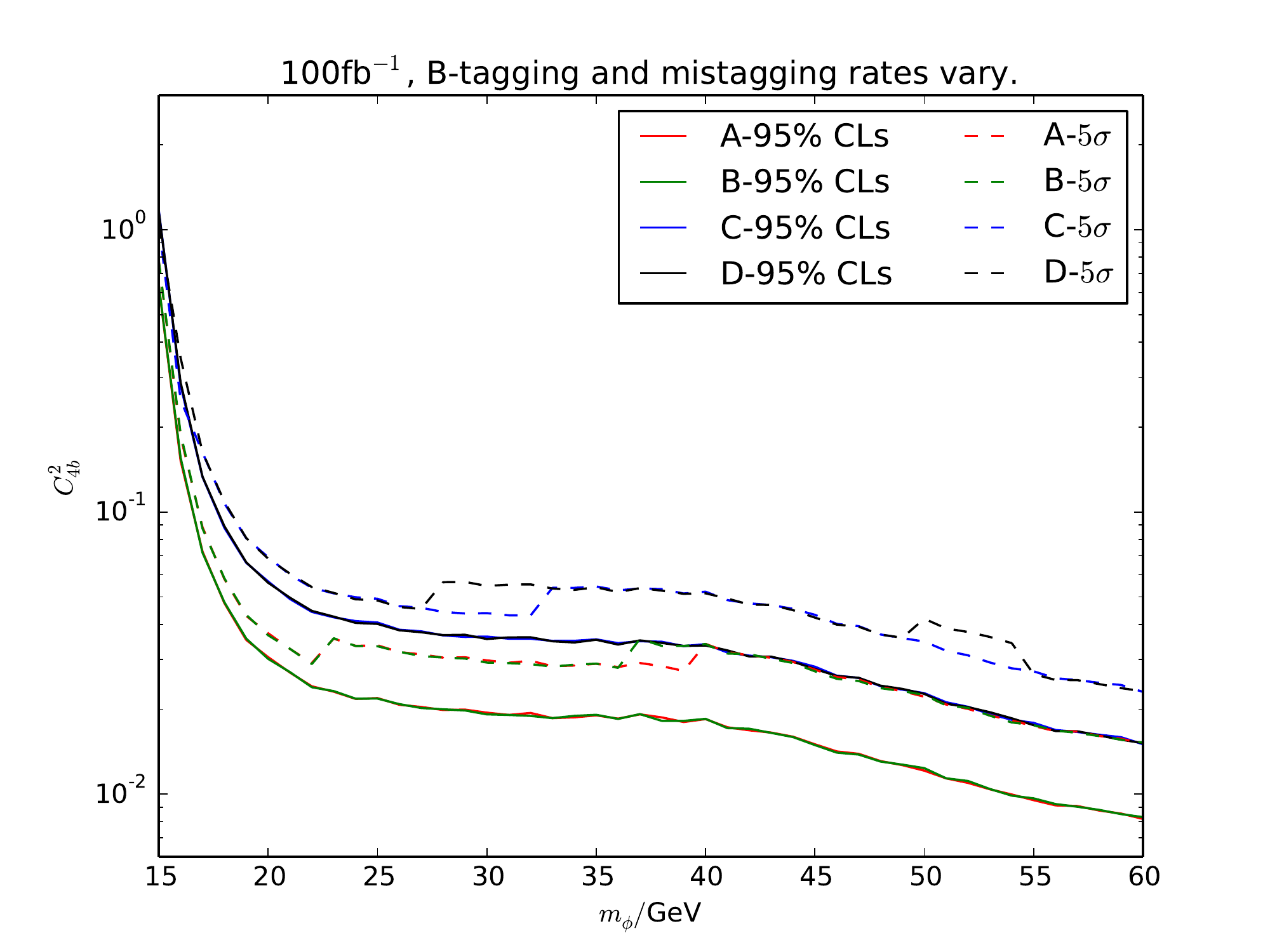}
\includegraphics[width=3.5in]{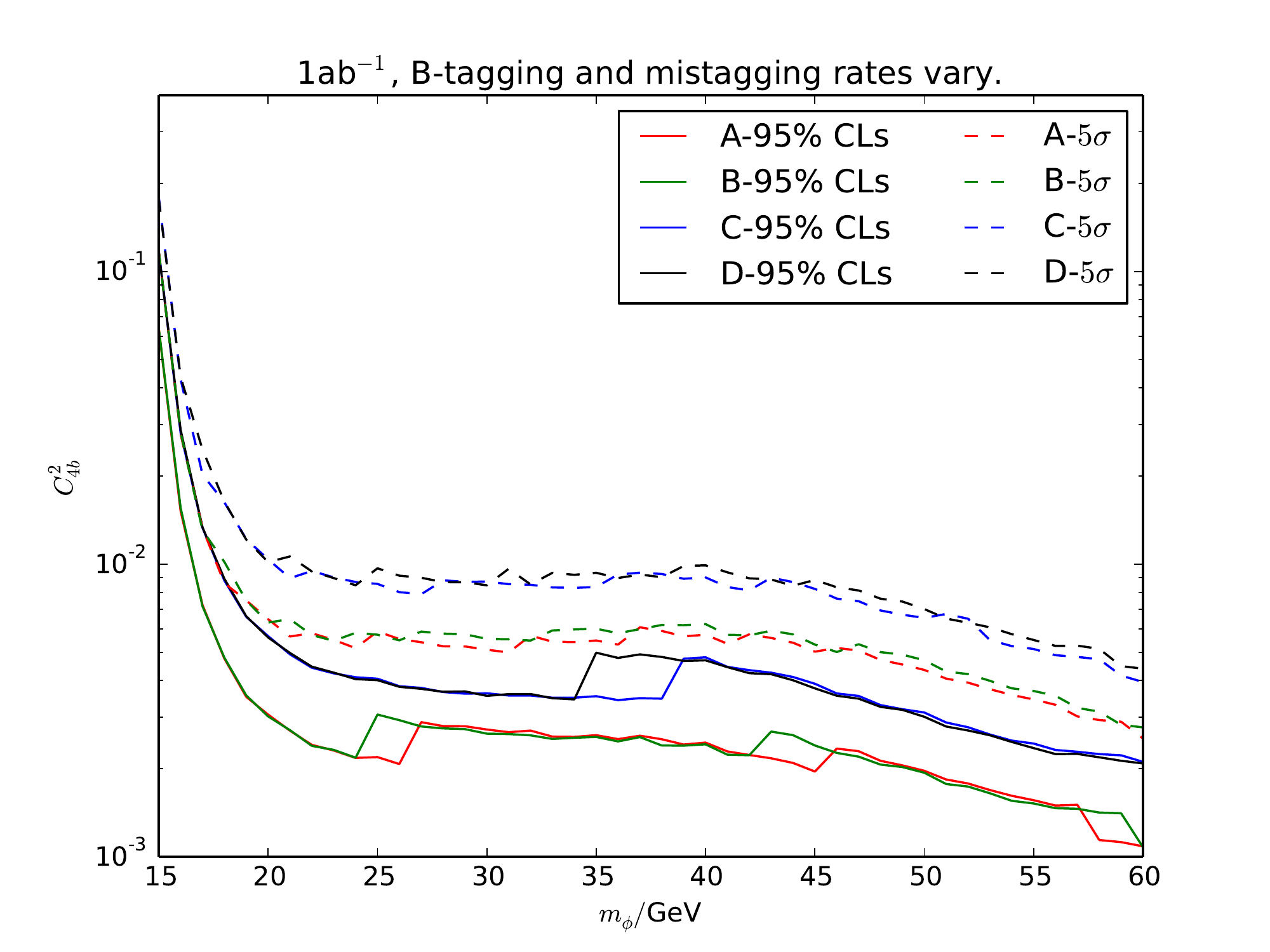}
\caption{\label{fig:cls1} Expected $95\%$ CLs exclusion limit (solid line) and
$5\sigma$ discovery reach (dashed line) in the ($C_{4b}^2,m_\phi$) plane at the LHeC.
Left: $100\fbi$ luminosity. Right: $1\abi$ luminosity. Different color
corresponds to different $b-$tagging scenarios
\eqref{eqn:ba}\eqref{eqn:bb}\eqref{eqn:bc}\eqref{eqn:bd} (see the text and legend).
$E_0=40\GeV$ is assumed.}
\end{figure*}

\begin{figure*}[ht]
\includegraphics[width=3.5in]{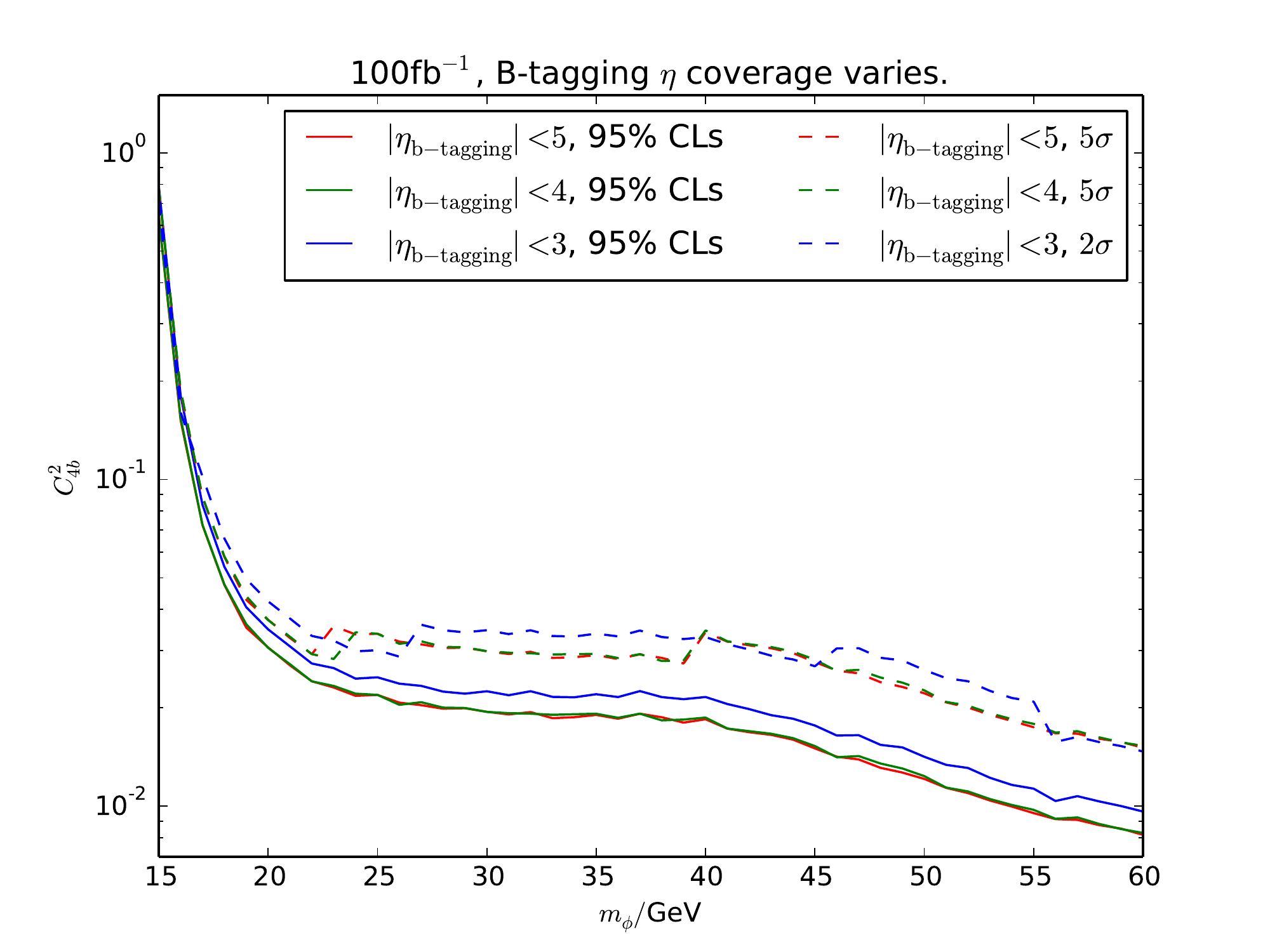}
\includegraphics[width=3.5in]{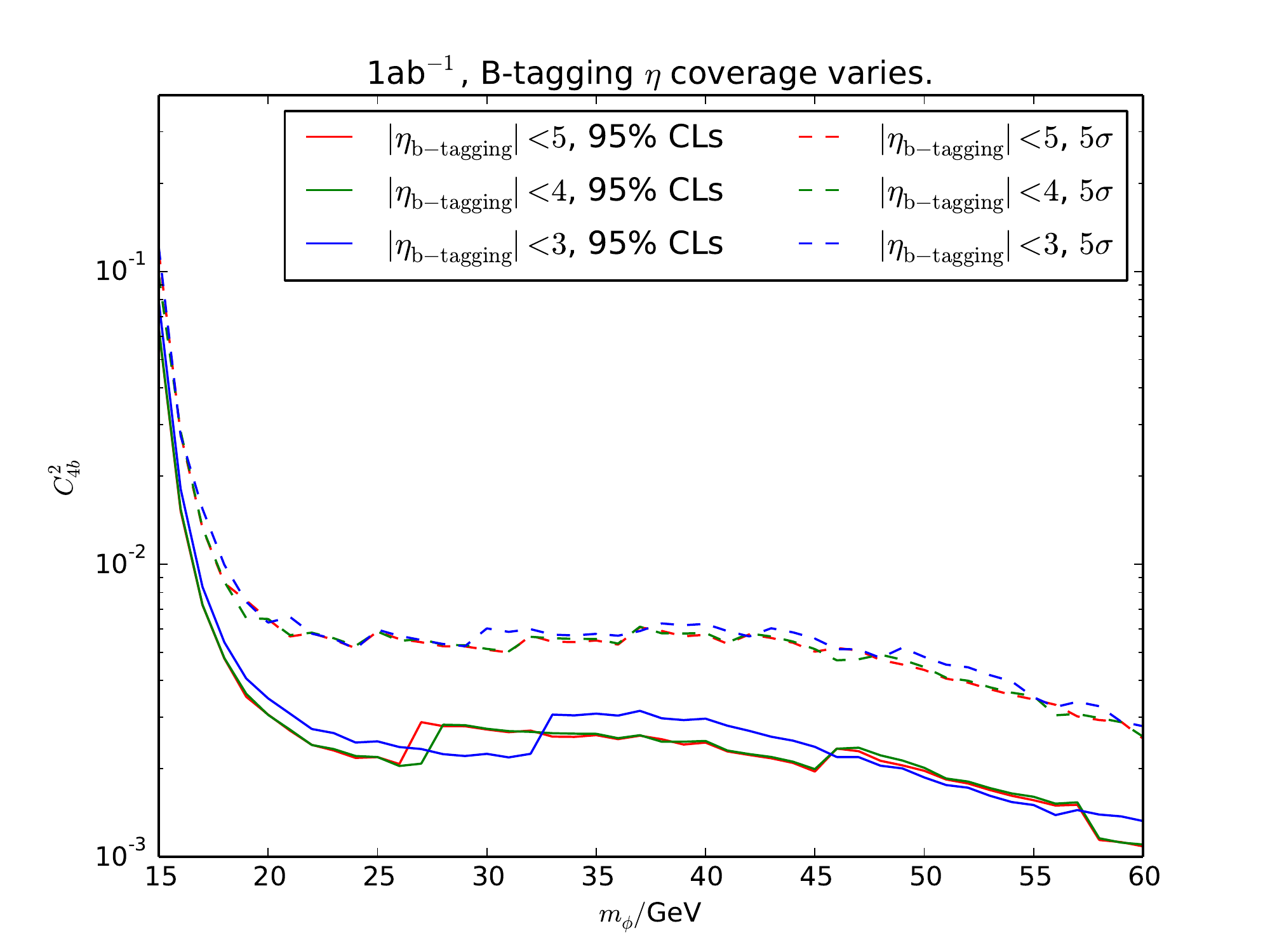}
\caption{\label{fig:cls2} Expected $95\%$ CLs exclusion limit (solid line) and
$5\sigma$ discovery reach (dashed line) in the ($C_{4b}^2,m_\phi$) plane at the LHeC.
Left: $100\fbi$ luminosity. Right: $1\abi$ luminosity. Different color
corresponds to different $b-$tagging pseudorapidity coverage (see the legend).
$E_0=40\GeV$ is assumed.}
\end{figure*}

We present cut flow tables (Table~\ref{table:cutflow20},
Table~\ref{table:cutflow40}, Table~\ref{table:cutflow60}) for three benchmark
masses $m_\phi=20,40,60\GeV$ under $b-$tagging performance scenario
\eqref{eqn:ba}. Only CC backgrounds are listed because PHP backgrounds
are expected to be negligible due to electron tagging and an
appropriate missing energy requirement. For the decay
of $t,W,Z$ in backgrounds, the following two cases are both considered and
included in our results. One is the decay to a minimal number of partons,
i.e. $t\to bqq, W\to qq,Z\to qq$, with each parton identified as one jet.
The other case is that one additional $b\bar{b}$ pair is radiated from
the decay products of $t,W,Z$. For $t+$jets the second kind of process
is found to contribute sizably to the total background. For the $h+$jets
background, only $h\to bb$ and $h\to 4b$ via tree-level SM processes are
considered. Due to limited Monte Carlo statistics, there are slight
differences among the three tables for the first four cuts on
backgrounds. The cross section numbers shown in the tables correspond to
the default choice $E_0=40\GeV$, except that in the last row of each table
for the signal and total background we show in parentheses the final
cross sections corresponding to $E_0=60\GeV$.
From the tables it can be concluded that the
$h\to\phi\phi\to 4b$ channel at the LHeC is almost background
free--with $100\fbi$ luminosity the expected number of background events is at
most $\ord(0.1)$ while the remaining signal cross section is
$\ord(1\fb)$ for $C_{4b}^2=1$. This is in sharp contrast to the
situation at the (HL-)LHC where the signal is buried in large top
quark backgrounds.

Fig.~\ref{fig:cls1} shows the expected
$95\%$ CLs~\cite{Read:2000ru,Read:2002hq} exclusion limits and
$5\sigma$ discovery reach at the LHeC for the $C_{4b}^2$ quantity in the
mass range $[15,60]\GeV$ assuming $100\fbi$ and $1\abi$ luminosity.
Various $b-$tagging performance scenarios are considered in the plots,
all assuming a $b-$tagging pseudorapidity coverage $|\eta|<5$.
Because the expected number of background events is quite small, in
setting exclusion limits and discovery reach we use exact formulae
of the Poisson distribution for a discrete random variable. This leads
to some small discontinuities at certain $m_\phi$ values when the
expected limits/reach are interpreted as the limits/reach for
the median of background-only or signal plus background hypothesis, as
can be seen from the plots.

From Fig.~\ref{fig:cls1} it can be easily seen that for $m_\phi$ in
the $[20,60]\GeV$ range the LHeC with
$100\fbi$ luminosity is capable of probing $C_{4b}^2$ to a few percent
level while with $1\abi$ luminosity the LHeC will eventually probe
$C_{4b}^2$ down to a few per mille level, both at $95\%$ CLs. We note that for
$m_\phi=20,40,60\GeV$, the $95\%$ CLs upper limit on $C_{4b}^2$ is about
$0.3\%(0.5\%),0.2\%(0.4\%),0.1\%(0.2\%)$ respectively, for $b-$tagging scenario \eqref{eqn:ba},
assuming $E_0=40\GeV(60\GeV)$. The result
is generally insensitive to mistag rates of $c-$jets, because with the
requirement of at least $4$ $b-$tagged jets, fake backgrounds do not
contribute much to the total background. On the other hand the final
signal rate is approximately proportional to the fourth power of the
$b-$tagging efficiency, thus it will be relatively important to
maintain a high $b-$tagging efficiency to retain more signal events.
As can be expected, the sensitivity drops quickly when $m_\phi$ becomes
smaller than about $20\GeV$ due to the collimation of $\phi$ decay
products that renders the resolved analysis inefficient. A jet
substructure analysis is needed to improve the sensitivity in this
mass region, which we leave for future study. On the other hand the
sensitivity improves as $m_\phi$ increases from about $40\GeV$ to
$60\GeV$. This is mainly because the $b-$jets from $h\rightarrow\phi\phi\rightarrow
4b$ decay in the $m_\phi=60\GeV$ case are more likely to pass
the basic cuts (especially, the $p_{Tj}>20\GeV$ cut) compared to
the $m_\phi=40\GeV$ case.

Fig.~\ref{fig:cls2} also shows the expected $95\%$ CLs exclusion limits and
$5\sigma$ discovery reach at the LHeC for the $C_{4b}^2$ quantity in the
mass range $[15,60]\GeV$ assuming $100\fbi$ and $1\abi$ luminosity. Here
$b-$tagging performance is fixed to scenario \eqref{eqn:ba} but various
$b-$tagging pseudorapidity coverage conditions are considered. The plots
indicate that the sensitivity reach of the LHeC for this channel is not
very sensitive to $b-$tagging pseudorapidity coverage.

\section{Constraints on the Higgs Singlet Extension of the SM}

\begin{figure*}[ht]
\begin{center}
\includegraphics[width=3.3in]{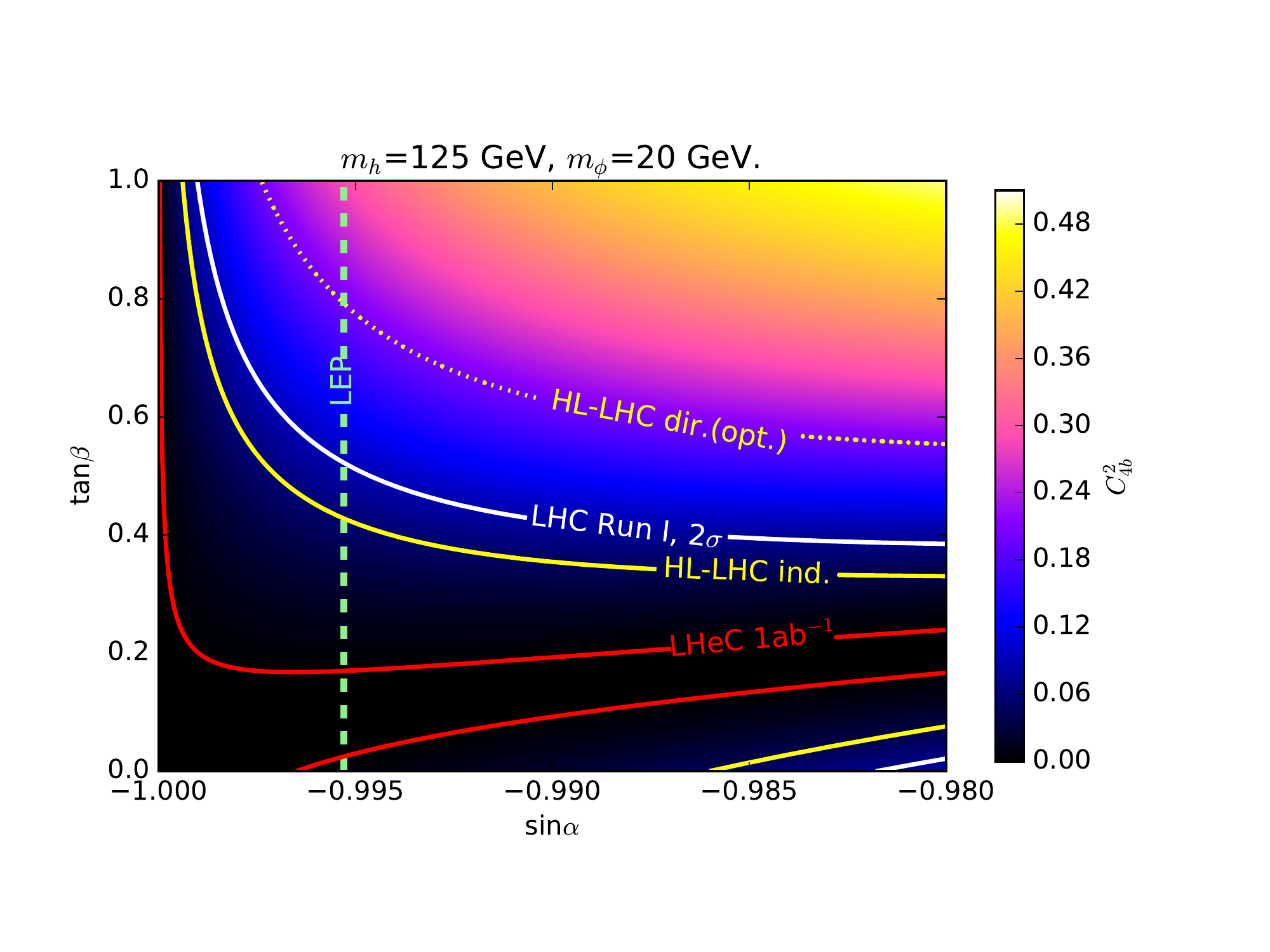}\\
\end{center}
\includegraphics[width=3.3in]{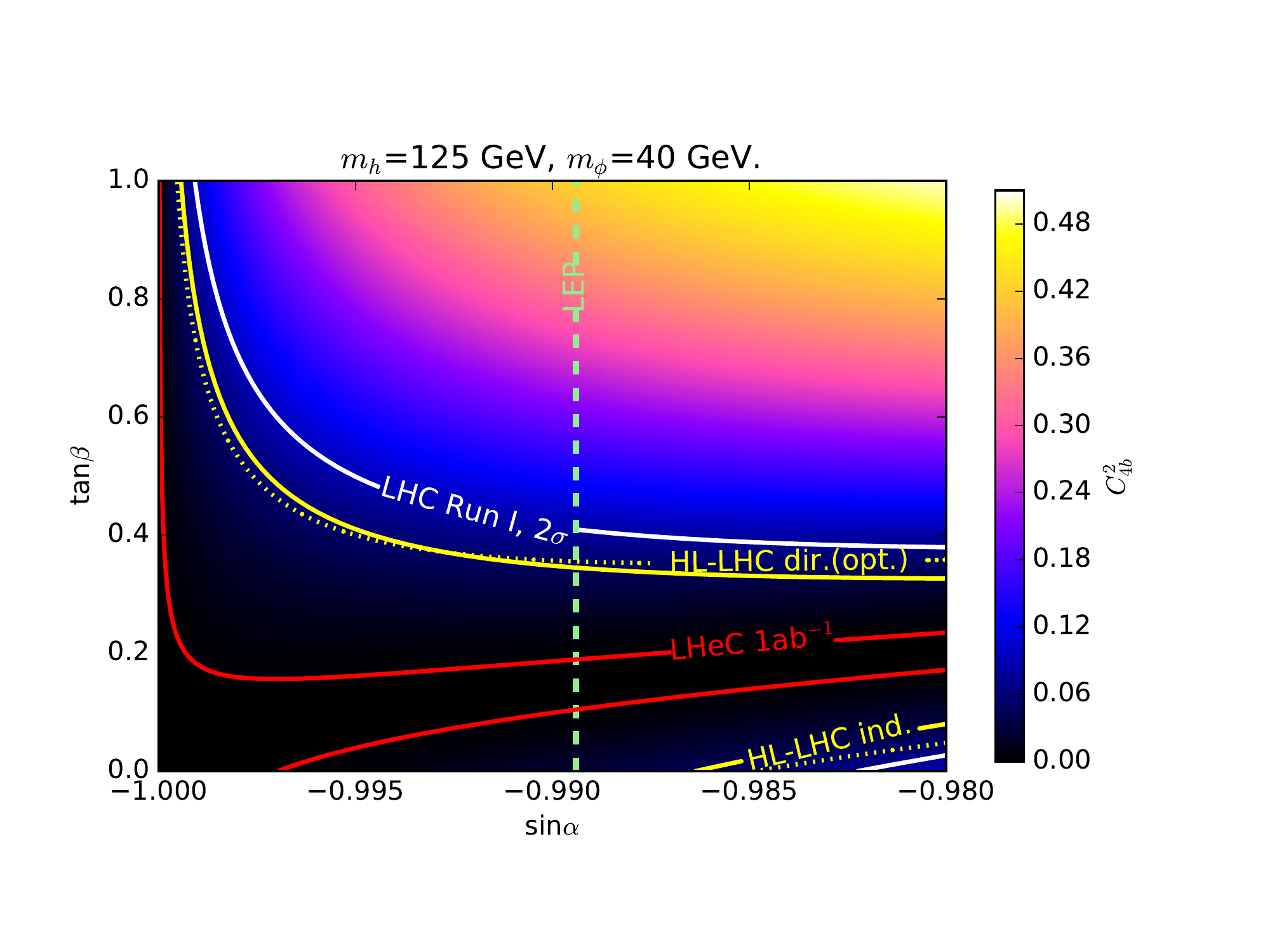}
\includegraphics[width=3.3in]{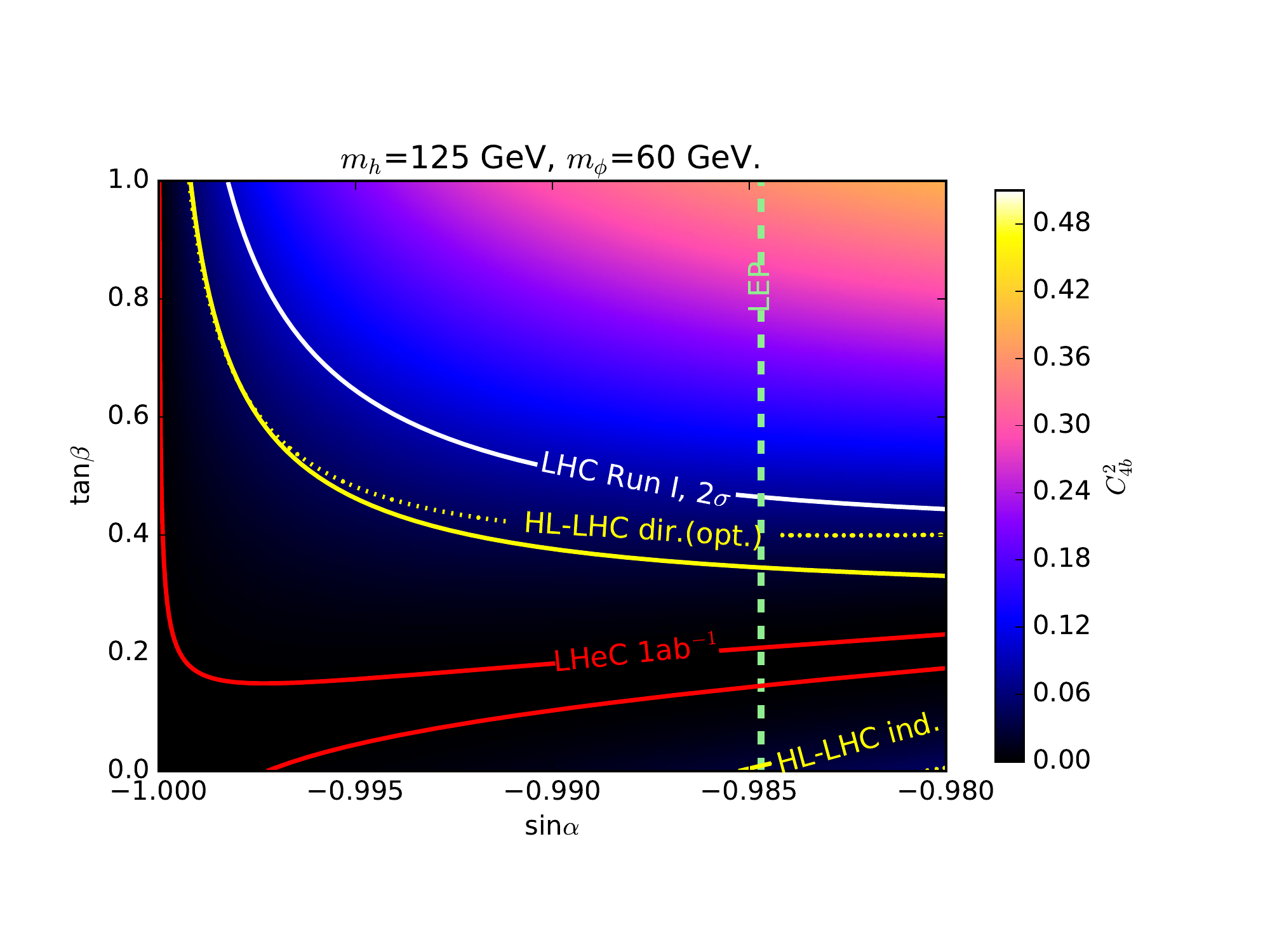}
\caption{\label{fig:hse} Future LHeC capability in probing the parameter
space of the Higgs singlet extension of the SM~\cite{Robens:2015gla},
plotted in the $\sin_\alpha-\tan_\beta$ plane for three benchmark
light Higgs masses $m_\phi=20,40,60\GeV$. Each point is colored
according to its $C_{4b}^2$ value for reference. Also shown
are current LEP and LHC bounds, and expected future HL-LHC
bounds. See text for detail.}
\end{figure*}

We now consider the interpretation of the expected sensitivity of
the LHeC in the context of Higgs singlet extension of the SM. For
simplicity we consider the Higgs singlet extension studied
in ~\cite{Robens:2015gla}. In this model, an additional real singlet
scalar $S$ is added to the SM. The Lagrangian of the Higgs kinetic
and potential terms is extended to the following form:
\begin{align}
\mathcal{L}_s=(D^{\mu}\Phi)^\dag D_\mu\Phi
+\partial^{\mu}S\partial_{\mu}S-V(\Phi,S)
\end{align}
with scalar potential
\begin{eqnarray}
 V(\Phi,S)&=&-m^2\Phi^{\dag}\Phi-\mu^2 S^2+\lambda_{1}(\Phi^{\dag}\Phi)^2 \nonumber \\
& & +\lambda_{2}S^4+\lambda_{3}\Phi^{\dag}\Phi S^2
\end{eqnarray}
Here $\Phi$ denotes the original SM Higgs doublet. The scalar potential
obeys a $Z_2$ symmetry. We allow $S$ to acquire vacuum expectation value
and express the Higgs fields in unitary gauge as
\begin{align}
\Phi\equiv\begin{pmatrix}0 \\ \frac{\tilde{h}+v}{\sqrt{2}}\end{pmatrix},
S\equiv\frac{h'+x}{\sqrt{2}}
\end{align}
Here $v=246\GeV$ ensures the correct mass generation for $W,Z$ bosons and
SM fermions. The gauge eigenstates $\tilde{h},h'$ can be related to mass
eigenstates $\phi,h$ via an orthogonal rotation
\begin{align}
\begin{pmatrix}\phi \\ h\end{pmatrix}
=\begin{pmatrix}\cos\alpha & -\sin\alpha \\ \sin\alpha & \cos\alpha\end{pmatrix}
\begin{pmatrix}\tilde{h} \\ h'\end{pmatrix}
\end{align}
Now it is convenient to parameterize the model in terms of five more
physical quantities: ($m_\phi,m_h$ are masses of $\phi$ and $h$ respectively)
\begin{align}
m_\phi,m_h,\alpha,v,\tan\beta\equiv\frac{v}{x}
\end{align}
The translation formulae between these quantities and original parameters
in the Lagrangian can be found in ~\cite{Robens:2015gla}. We are interested
in the case in which the additional Higgs boson is lighter, therefore we
fix $m_h=125\GeV$ and allow three parameters $m_\phi,\alpha,\tan\beta$ to vary.
Here we focus on the more interesting region where $\sin\alpha\rightarrow-1$
which also allows for a special direction $\tan\beta=-\cot\alpha$ that results
in a vanishing $\Br(h\to\phi\phi)$~\cite{Robens:2015gla}. We consider three
benchmark values of \linebreak[4] $m_\phi$ ($m_\phi=20,40,60\GeV$) and plot the current LEP
and LHC constraints and future HL-LHC and LHeC constraints on the
$\tan\beta-\sin\alpha$ plane, see Fig.~\ref{fig:hse}. Each point is colored
according to its $C_{4b}^2$ value for reference. The factor
$\Br^2(\phi\to b\bar{b})$ appeared in $C_{4b}^2$ definition Eq.~\eqref{eqn:c4b2}
is almost at a constant value $0.77$ in the mass range $m_\phi\in[20,60]\GeV$
~\cite{Curtin:2014pda}. The deep black regions which corresponds to very small
$C_{4b}^2$ values slightly tilt upwards with the decreasing of $|\sin\alpha|$
from about $0.995$ to $0.980$. In this range these regions just center around
the abovementioned special direction $\tan\beta=-\cot\alpha$ which makes
$\Br(h\to\phi\phi)$ vanish~\cite{Robens:2015gla} and renders its vicinity
difficult to probe through exotic Higgs decay search.
The LEP constraints (green dashed line) come from direct
search for additional Higgs bosons and is taken directly from
~\cite{Robens:2015gla}. Points on the right side of the green dashed line
is excluded at $95\%$ confidence level. This indicates that LEP search forces
the mixing between two Higgses to be very small for the scenario in which
there is a light Higgs boson in the mass range ($m_\phi\in[20,60]\GeV$). In such
a case there cannot be sizable deviation of Higgs signal strength due to Higgs
mixing. However, the opening of exotic Higgs decay $h\to\phi\phi$ could
lead to siazble suppression of $125\GeV$ Higgs signal strengths. The LHC
Run I constraints (white solid line) come from the $125\GeV$ Higgs signal strength
measurements~\cite{Khachatryan:2016vau}. The regions between the two white solid
lines for $m_\phi=20,40\GeV$ (and the region below the white solid lines for
$m_\phi=60\GeV$ case) are allowed by LHC Run I measurements at $2\sigma$
level. We translated the HL-LHC projection of the precision of Higgs signal
strength measurements~\cite{ATL-PHYS-PUB-2014-016} into constraints (yellow
solid line "HL-LHC ind.") on the parameter space of the Higgs singlet
extension of the SM (assuming half theoretical uncertainties, according
to ~\cite{ATL-PHYS-PUB-2014-016}). At the (HL-)LHC, $h\to\phi\phi\to 4b$
can be directly probed via $Wh$ associated production, as has been done by
ATLAS~\cite{Aaboud:2016oyb}. However, current constraint from this method
is quite weak and even $C_{4b}^2=1$ cannot be bound. We extrapolate the
current constraint~\cite{Aaboud:2016oyb} to $3\abi$ HL-LHC, with a very
optimistic assumption that all selection efficiency can be maintained
and all systematic uncertainties scale with the square root of luminosity.
The corresponding $95\%$ CLs exclusion limits is plotted as yellow
dotted line "HL-LHC dir.(opt.)". It can be seen that even with this very
optimistic assumption the sensitivity of the direct search from $Wh$
channel is at most comparable to the indirect constraint from the HL-LHC
$125\GeV$ Higgs signal strength measurements. The LHeC $1\abi$ $95\%$ CLs
sensitivity is plotted as the red solid lines, assuming $b-$tagging
scenario \eqref{eqn:ba}, $b-$tagging pseudorapidity coverage $|\eta|<5.0$
and $E_0=40\GeV$. The LHeC is expected to exclude region outside the red
solid lines if no new physics exists. It is obvious that the LHeC
exclusion capability extends to the deep black region which represents
very small $C_{4b}^2$ values. If no lepton colliders are available before
the end of the HL-LHC, much of the parameter space of the Higgs singlet
extension model could only be reached via the $ep$ machine.

\section{Discussion and Conclusion}
In this paper we studied the LHeC sensitivity to the exotic Higgs decay
process $h\to\phi\phi\to 4b$ in which $\phi$ denotes a spin-0 particle
lighter than half of $125\GeV$. We performed a parton level analysis
and showed that with $1\abi$ luminosity the LHeC is able to exclude
$C_{4b}^2$ at a fer per mille level ($95\%$ CLs), when only statistical uncertities
are included. To maintain the sensitivity, it is important to choose
a $b-$tagging working point with relatively large $b-$tagging efficiency.
The sensitivity is not very sensitive to the variation of $b-$tagging
pseudorapidity coverage from $3$ to $5$. Using the Higgs singlet
extension of the SM as an illustration, we showed
that the LHeC direct search of $h\to\phi\phi\to 4b$ is the most sensitive
probe of much of the parameter space of the model in the future, if no
lepton colliders are available. Of course this LHeC search will also
deliver significant impacts on the scalar sector of other BSM theories
when one of the scalar boson lies in the mass range $\sim(2m_b,m_h/2)$.

The analysis presented here can be further improved in several aspects.
First is of course a more realistic estimation of the signal and
backgrounds including parton shower and more detailed detector effects.
Especially for multijet final states a parton shower correctly merged
to matrix element will be highly desirable. Secondly, we could further
utilize the sample with the requirement of less $b-$tagged jets or even
less reconstructed jets, e.g. three $b-$tagged jets. This technique has
already been used in ~\cite{Aaboud:2016oyb} and is expected to further
improve the sensitivity, especially in the first stages of data
collection when statistics is small. Thirdly, we have only applied a
cut-based analysis with very simple variables. A further multivariate
analysis may deliver additional gain in sensitivity. Furthermore,
the sensitivity in the $m_\phi<20\GeV$ mass range could be improved
via a jet substructure analysis, as has been emphasized. Besides these
directions of exploration, it should however be emphasized that in the
current analysis PHP backgrounds are assumed to be negligible
compared to CC backgrounds under the condition discussed in
Section ~\ref{sec:cs}. A more detailed detector simulation is thus
needed to pin down the event selection conditions required to suppress
PHP backgrounds. We also note that in the present study systematic
uncertainties have not been included. However, since the expected
background event number is very small, we expect that the obtained
sensitivity (discovery and exclusion reach) would be qualitatively
stable against systematic uncertainties, which means that the $1\abi$
LHeC could still do much better than the HL-LHC with respect to
the $h\rightarrow\phi\phi\rightarrow 4b$ search.

The exotic Higgs decays constitute an intriguing and important part of
Higgs physics which deserve comprehensive theoretical and experimental
investigations. Previous attempts and attention have nearly all been
devoted to hadron-hadron collisions or $e^{+}e^{-}$ collisions.
We demonstrate in this paper that for certain important processes
which suffer from large backgrounds in hadron-hadron collisions, it is
clearly superior to conduct the search at a concurrent $ep$ collider,
if an $e^{+}e^{-}$ machine with sufficient center-of-mass energy
is not available. In that case, it is highly expected that the $ep$
machine will play an important role in precision Higgs
studies, including the study of exotic Higgs decays like
$h\to\slashed{E}_T$~\cite{Tang:2015uha}, $h\to\phi\phi\to 4b$ and other
channels beset by jets or
$\slashed{E}_T$~\cite{AristizabalSierra:2008ye,Huang:2013ima}.

%

\begin{acknowledgements}
We would like to thank Qing-Hong Cao, Manuel Drees, Uta Klein,
Masahiro Kuze, Yan-Dong Liu, Ying-Nan Mao, Masahiro Tanaka and
Hao Zhang for helpful discussions. This work was supported in part by
the Natural Science Foundation of China (Grants No. 11135003,
No. 11375014 and No. 11635001) and by the China Postdoctoral
Science Foundation under Grant No. 2016M600006.
\end{acknowledgements}

\bibliographystyle{spphys}       
\bibliography{lhech24b-epjc}   

\begin{thebibliography}{10}
\providecommand{\url}[1]{{#1}}
\providecommand{\urlprefix}{URL }
\expandafter\ifx\csname urlstyle\endcsname\relax
  \providecommand{\doi}[1]{DOI \discretionary{}{}{}#1}\else
  \providecommand{\doi}{DOI \discretionary{}{}{}\begingroup
  \urlstyle{rm}\Url}\fi

\bibitem{Chatrchyan:2012ufa}
S.~Chatrchyan, et~al., Phys.Lett. \textbf{B716}, 30 (2012).
\newblock \doi{10.1016/j.physletb.2012.08.021}

\bibitem{Aad:2012tfa}
G.~Aad, et~al., Phys.Lett. \textbf{B716}, 1 (2012).
\newblock \doi{10.1016/j.physletb.2012.08.020}

\bibitem{Kling:2016opi}
F.~Kling, J.M. No, S.~Su, JHEP \textbf{09}, 093 (2016).
\newblock \doi{10.1007/JHEP09(2016)093}

\bibitem{Khachatryan:2016vau}
G.~Aad, et~al.,   (2016)

\bibitem{Curtin:2013fra}
D.~Curtin, R.~Essig, S.~Gori, P.~Jaiswal, A.~Katz, et~al., Phys.Rev.
  \textbf{D90}(7), 075004 (2014).
\newblock \doi{10.1103/PhysRevD.90.075004}

\bibitem{Mao:2016jor}
Y.n. Mao, S.h. Zhu,   (2016)

\bibitem{Cao:2013gba}
J.~Cao, F.~Ding, C.~Han, J.M. Yang, J.~Zhu, JHEP \textbf{1311}, 018 (2013).
\newblock \doi{10.1007/JHEP11(2013)018}

\bibitem{Cheung:2007sva}
K.~Cheung, J.~Song, Q.S. Yan, Phys.Rev.Lett. \textbf{99}, 031801 (2007).
\newblock \doi{10.1103/PhysRevLett.99.031801}

\bibitem{Carena:2007jk}
M.~Carena, T.~Han, G.Y. Huang, C.E. Wagner, JHEP \textbf{0804}, 092 (2008).
\newblock \doi{10.1088/1126-6708/2008/04/092}

\bibitem{Kaplan:2011vf}
D.E. Kaplan, M.~McEvoy, Phys.Rev. \textbf{D83}, 115004 (2011).
\newblock \doi{10.1103/PhysRevD.83.115004}

\bibitem{Kaplan:2009qt}
D.E. Kaplan, M.~McEvoy, Phys.Lett. \textbf{B701}, 70 (2011).
\newblock \doi{10.1016/j.physletb.2011.05.026}

\bibitem{Aaboud:2016oyb}
M.~Aaboud, et~al.,   (2016)

\bibitem{AbelleiraFernandez:2012cc}
J.~Abelleira~Fernandez, et~al., J.Phys. \textbf{G39}, 075001 (2012).
\newblock \doi{10.1088/0954-3899/39/7/075001}

\bibitem{Bruening:2013bga}
O.~Bruening, M.~Klein, Mod. Phys. Lett. \textbf{A28}(16), 1330011 (2013).
\newblock \doi{10.1142/S0217732313300115}

\bibitem{Jarlskog:1990dv}
G.~Jarlskog, D.~Rein (eds.).
\newblock \emph{{ECFA Large Hadron Collider Workshop, Aachen, Germany, 4-9 Oct
  1990: Proceedings.1.}} (1990).
\newblock \urlprefix\url{http://weblib.cern.ch/abstract?90-10}

\bibitem{Han:2009pe}
T.~Han, B.~Mellado, Phys.Rev. \textbf{D82}, 016009 (2010).
\newblock \doi{10.1103/PhysRevD.82.016009}

\bibitem{Biswal:2012mp}
S.S. Biswal, R.M. Godbole, B.~Mellado, S.~Raychaudhuri, Phys.Rev.Lett.
  \textbf{109}, 261801 (2012).
\newblock \doi{10.1103/PhysRevLett.109.261801}

\bibitem{Cakir:2013bxa}
I.~Cakir, O.~Cakir, A.~Senol, A.~Tasci, Mod.Phys.Lett. \textbf{A28}(31),
  1350142 (2013).
\newblock \doi{10.1142/S0217732313501423}

\bibitem{Senol:2012fc}
A.~Senol, Nucl.Phys. \textbf{B873}, 293 (2013).
\newblock \doi{10.1016/j.nuclphysb.2013.04.016}

\bibitem{Tang:2015uha}
Y.L. Tang, C.~Zhang, S.h. Zhu, Phys. Rev. \textbf{D94}(1), 011702 (2016).
\newblock \doi{10.1103/PhysRevD.94.011702}

\bibitem{Zhe:2011yr}
W.~Zhe, W.~Shao-Ming, M.~Wen-Gan, G.~Lei, Z.~Ren-You, Phys.Rev. \textbf{D83},
  055003 (2011).
\newblock \doi{10.1103/PhysRevD.83.055003}

\bibitem{Klein:2015hcc}
U.~Klein.
\newblock Higgs heavy flavour decay studies using jet probabilities.
\newblock Talk given at LHeC Workshop 2015

\bibitem{Kumar:2015tua}
M.~Kumar, X.~Ruan, A.S. Cornell, R.~Islam, B.~Mellado, J.Phys.Conf.Ser.
  \textbf{623}(1), 012017 (2015).
\newblock \doi{10.1088/1742-6596/623/1/012017}

\bibitem{Kumar:2015kca}
M.~Kumar, X.~Ruan, R.~Islam, A.S. Cornell, M.~Klein, U.~Klein, B.~Mellado,
  (2015)

\bibitem{Alloul:2013bka}
A.~Alloul, N.D. Christensen, C.~Degrande, C.~Duhr, B.~Fuks, Comput.Phys.Commun.
  \textbf{185}, 2250 (2014).
\newblock \doi{10.1016/j.cpc.2014.04.012}

\bibitem{Alwall:2014hca}
J.~Alwall, R.~Frederix, S.~Frixione, V.~Hirschi, F.~Maltoni, et~al., JHEP
  \textbf{1407}, 079 (2014).
\newblock \doi{10.1007/JHEP07(2014)079}

\bibitem{Ball:2013hta}
R.D. Ball, V.~Bertone, S.~Carrazza, L.~Del~Debbio, S.~Forte, A.~Guffanti, N.P.
  Hartland, J.~Rojo, Nucl. Phys. \textbf{B877}, 290 (2013).
\newblock \doi{10.1016/j.nuclphysb.2013.10.010}

\bibitem{Jager:2010zm}
B.~Jager, Phys.Rev. \textbf{D81}, 054018 (2010).
\newblock \doi{10.1103/PhysRevD.81.054018}

\bibitem{Gaddi:2015det}
A.~Gaddi.
\newblock Lhec detector: Preliminary engineering study.
\newblock talk given at LHeC Workshop 2015

\bibitem{Conte:2012fm}
E.~Conte, B.~Fuks, G.~Serret, Comput.Phys.Commun. \textbf{184}, 222 (2013).
\newblock \doi{10.1016/j.cpc.2012.09.009}

\bibitem{Klasen:1998cw}
M.~Klasen, Eur. Phys. J. \textbf{C7}, 225 (1999).
\newblock \doi{10.1007/s100529801006}

\bibitem{Adloff:1999ah}
C.~Adloff, et~al., Eur. Phys. J. \textbf{C13}, 609 (2000).
\newblock \doi{10.1007/s100520050721}

\bibitem{Read:2000ru}
A.L. Read, in \emph{{Workshop on confidence limits, CERN, Geneva, Switzerland,
  17-18 Jan 2000: Proceedings}} (2000), pp. 81--101.
\newblock \urlprefix\url{http://weblib.cern.ch/abstract?CERN-OPEN-2000-205}

\bibitem{Read:2002hq}
A.L. Read, J. Phys. \textbf{G28}, 2693 (2002).
\newblock \doi{10.1088/0954-3899/28/10/313}.
\newblock [,11(2002)]

\bibitem{Robens:2015gla}
T.~Robens, T.~Stefaniak, Eur. Phys. J. \textbf{C75}, 104 (2015).
\newblock \doi{10.1140/epjc/s10052-015-3323-y}

\bibitem{Curtin:2014pda}
D.~Curtin, R.~Essig, Y.M. Zhong, JHEP \textbf{06}, 025 (2015).
\newblock \doi{10.1007/JHEP06(2015)025}

\bibitem{ATL-PHYS-PUB-2014-016}
{Projections for measurements of Higgs boson signal strengths and coupling
  parameters with the ATLAS detector at a HL-LHC}.
\newblock Tech. Rep. ATL-PHYS-PUB-2014-016, CERN, Geneva (2014).
\newblock \urlprefix\url{http://cds.cern.ch/record/1956710}

\bibitem{AristizabalSierra:2008ye}
D.~Aristizabal~Sierra, W.~Porod, D.~Restrepo, C.E. Yaguna, Phys. Rev.
  \textbf{D78}, 015015 (2008).
\newblock \doi{10.1103/PhysRevD.78.015015}

\bibitem{Huang:2013ima}
J.~Huang, T.~Liu, L.T. Wang, F.~Yu, Phys. Rev. Lett. \textbf{112}(22), 221803
  (2014).
\newblock \doi{10.1103/PhysRevLett.112.221803}

\end{thebibliography}


\end{document}